\documentclass[Crown,sageh,times]{sagej}

\usepackage{moreverb,url}

\usepackage[colorlinks,bookmarksopen,bookmarksnumbered,citecolor=red,urlcolor=red]{hyperref}

\usepackage{latexsym}
\usepackage{amsmath}
\usepackage{xr}
\usepackage{moreverb}
\usepackage{graphicx}
\usepackage{blindtext}
\usepackage{xcolor}
\usepackage{hyperref}

\newcommand\BibTeX{{\rmfamily B\kern-.05em \textsc{i\kern-.025em b}\kern-.08em
T\kern-.1667em\lower.7ex\hbox{E}\kern-.125emX}}

\usepackage[font={bf, scriptsize}]{subcaption}
\usepackage{xspace}
\begin{document}

\runninghead{Wu, Li and Feng}

\title{Z-residual diagnostics for detecting misspecification of the functional form of covariates for shared frailty models
%A demonstration of the \LaTeXe\ class file for\itshape{SAGE Publications}
}

\author{Tingxuan Wu \affilnum{1} and  Longhai Li \affilnum{2} and  Cindy Feng \affilnum{3}}

\affiliation{\affilnum{1,2}Department of Mathematics and Statistics, University of Saskatchewan, Saskatoon, CA\\
\affilnum{3}Department of Community Health and Epidemiology, Faculty of Medicine, Dalhousie University, Halifax, CA}

\corrauth{Cindy Feng, Department of Community Health and Epidemiology, Faculty of Medicine, Dalhousie University, Halifax, CA.}

\email{cindy.feng@dal.ca}

\begin{abstract}
%This paper describes the use of the \LaTeXe\
%\textsf{\journalclass} class file for setting papers to be
%submitted to a \textit{SAGE Publications} journal.
%The template can be downloaded \href{http://www.uk.sagepub.com/repository/binaries/SAGE LaTeX template.zip}{here}.
% !TEX root = simdoc.tex

In survival analysis, the hazard function often depends on a set of covariates.  Martingale and deviance residual are most widely used for examining the validity of  the function form of covariates by checking whether there is a discernible trend in  their scatterplot against continuous covariates. However, visual inspection of martingale and deviance residuals is often subjective. In addition, these residuals lack a reference distribution due to censoring. It is therefore challenging to derive numerical statistical tests based on martingale or deviance residuals. In this paper, we extend  the idea of randomized survival probability (Li et al. 2021)  and develop a residual diagnostic tool that can provide both graphical and numerical tests for checking the covariate functional form in semi-parametric shared frailty models. We develop  a general function that calculates Z-residuals for semi-parametric shared frailty models based on the output from the \texttt{coxph} function in the \texttt{survival} package in R. Our extensive simulation studies indicate  that the derived numerical test based on Z-residuals has great  power for checking the functional form of covariates. In a real data application on modelling the survival time of acute myeloid leukemia patients, the Z-residual diagnosis results show  that a model with log-transformation is inappropriate for modelling the survival time, which could not be detected by other diagnostic methods.
%and proposed a non-homogeneity test for testing whether there is a trend in Z-residuals

\end{abstract}


\maketitle

% !TEX root = simdoc.tex

% new command
\newcommand \tistar {t^{*}_{ij}}
\newcommand \Tistar {T^{*}_{ij}}
\newcommand \SiTistar {S_{ij}(\Tistar)}
\newcommand \rcTistar {r^c_{ij} (\Tistar)}
\newcommand \SP{survival probability\xspace}
\newcommand \SPs{survival probabilities\xspace}
\newcommand\nrsp{\mbox{\scriptsize NRSP}}
\newcommand\rsp{S_{ij}^{R}(T_{ij}, d_{ij}, U_{ij})}
\newcommand \nref {[\textbf{references}]\xspace}
\newcommand\svf{S_{ij}(\cdot)}
\newcommand\ques{\textbf{[questions]}}
\newcommand\pmin{p_{\mbox{\scriptsize min}}}
\newcommand\obs{^{\scriptsize \mbox{obs}}}

%Detecting misspecification of the functional form of covariate effect for shared frailty models using Z-residuals

\newpage
\section{Introduction}

Survival data with a multilevel structure occur frequently in many applications. For example, patients are often clustered within hospitals. The hazard of events differs from one cluster to another cluster induced by unobserved cluster-level factors. In survival analysis, conventional Cox proportional hazard models \citep{CoxD.R.1972RMaL} and accelerated failure time models \citep{LIND.Y.1998Aftm} assume that subjects are independent. Random effects can be incorporated into conventional survival models to account for cluster-level heterogeneity. Such heterogeneity is often called frailty in the context of survival analysis. A shared frailty model extends the classic survival models by incorporating random effects (frailties) acting multiplicatively on the baseline hazard function \citep{vaupel_impact_1979}, where the frailties are common or shared among individuals within a cluster or group \citep{clayton_model_2022,duchateau_frailty_2008,KaragrigoriouAlex2011FMiS,HanagalDD2015Msdu}.  Despite the increasing popularity of shared frailty models for modeling clustered survival data,  examining model assumptions are often overlooked partly due to the limited model diagnostic tools. % For instance,  we often assume that continuous covariates have a linear form shared frailty models. However, this assumption should be checked. %Despite the increasing applications of shared frailty models in medical research and other fields, the validation of the correctness of the fitted shared frailty model for a given dataset is still often neglected in practice. Therefore, the choice of the model for a dataset is often subjective without undergoing serious scrutiny. 

Residual diagnostics are often used to assess the overall goodness of fit (GOF) and to identify specific model misspecification (e.g., functional form of covariate effects). However, in the presence of censored observations, residual diagnostics is not as straightforward as a normal linear regression model. Cox-Snell (CS) residual \citep{CoxD.R.1968AGDo} is the most widely used tool for diagnosing survival models, which are defined as the negative logarithm of estimated survival probability. In the absence of censored observations, the survival probability is uniformly distributed when the model is true; therefore, the CS residual is exponentially distributed. However, in the presence of censored observations, CS residuals are no longer exponentially distributed since the survival probability is not uniformly distributed. To account for censored observations, diagnostics based on CS residuals compares the agreement of cumulative hazard plot of CS residuals estimated with Kaplan-Meier method \citep{kaplan_nonparametric_1958} and the $45^\circ$ straight line, which is the cumulative hazard of the standard exponential distribution.

Although the overall GOF checking such as the cumulative hazard plot of CS residuals is widely used for diagnosing survival models, the overall GOF test reveals little information about the nature of the model inadequacies. Tailored graphical and numeric diagnostic tools are therefore needed. A number of residuals diagnostics tools have been proposed \citep{collett_modelling_2015} for checking the functional form of covariates, of which martingale  \citep{therneau_martingale-based_1990} and deviance \citep{therneau_modeling_2013,mccullagh_generalized_1989} residuals are most widely used. Martingale residuals can be viewed as the difference between the observed value of a subject’s failure indicator and its expected value, integrated over the time for which that patient was at risk, which can be used to assess the functional form covariates and identify outliers in the survival data. 
Deviance residuals are a normalized transform of the martingale residuals. They also have a mean of zero but are approximately symmetrically distributed about zero when the fitted model is appropriate. Although these two types of residuals are widely used and available in the \texttt{survival} package in R software, each of these traditional types of residuals has limitations. Martingale residuals are  asymmetric, with the upper bound of martingale residuals being one and no lower bound, making it difficult for visual inspection.  Deviance residuals are less skewed and more normally distributed. The locally weighted scatterplot smoothing (LOWESS) lines  on the scatterplots of the residuals against the continuous covariates is useful for  revealing patterns in the residuals that would not otherwise be perceived. However, visual inspection of LOWESS lines can be still subjective.  It is desirable to have a numerical measure of the statistical  significance of the observed trend. However, martingale and deviance residuals lack a reference distribution due to censoring. It is therefore challenging to derive a numerical test to measure the statistical significance of the observed pattern in the residual plots.

\citet{LiLonghai2021Mdfc} proposed to use randomized survival probabilities (RSPs) to define residuals for checking the model assumptions of accelerated failure time (AFT) models without random effects. The key idea of RSP is to replace the survival probability of a censored failure time with a uniform random number between 0 and the survival probability of the censored time. The RSPs are uniformly distributed under the true model, hence, can then be transformed into normally distributed residuals with the normal quantile function. The new residual was called the normally-transformed RSP (NRSP) residual.  Provided with the normally distributed reference distribution for the NRSP residual,  statistical tests can be derived based on NRSP residuals for checking model assumptions, such as distributional assumption, functional form of covariates, etc. However, NRSP residuals have not been extended to diagnose Cox proportional hazard models or semi-parametric shared frailty models. %It is also unclear if the overall model diagnosis methods based on this type of new residual diagnosis methods could adequately capture the misspecifcation of functional form of covariates. 

In this study, we extend the idea of NRSP residuals to develop residual diagnostics tools for checking the functional form of the covariates in semi-parametric shared frailty models. We rename NRSP residuals as Z-residuals for simplicity, as Z is often used to denote a standard normal random variable. For calculating the Z-residuals, we treat the random effects as fixed effects; that is, our Z-residual is conditional on the group identities. We developed a general function for calculating such conditional Z-residuals given the output of \texttt{coxph} in the \texttt{survival} package in R and proposed a non-homogeneity test for testing whether there is a trend in Z-residuals.  We conducted extensive  simulation studies to investigate the performance of the Z-residuals diagnostics tool in detecting misspecification of functional form of covariates. Our results showed that the non-homogeneity test based on Z-residuals has greater power  and satisfactory type I error compared to the overall GOF tests in detecting misspecification of the covariate functional form. We also demonstrated the effectiveness of Z-residuals in diagnosing functional form of covariates in a real data analysis of mortality risk of acute myeloid leukemia patients \citep{ESTEYE.H2000Eott,HendersonRobin2002MSVi}. Our proposed Z-residual diagnostic tool discovered that a model with log transformation of a continuous covariate is inappropriate in this real data application, which however can not be captured by other diagnostic methods.

The rest of this paper is organized as follows. Section \ref{sec:model} gives a brief review of semi-parametric shared frailty models. In Section \ref{sec:review} we review the conventional residuals and model diagnostics methods for shared frailty models. In Section \ref{sec:zresidual} we present the definition of Z-residuals and the non-homogeneity test based on Z-residuals. In Section \ref{sec:nl}, we conduct simulation studies to investigate the performances of the Z-residual diagnostics  tool. Section \ref{sec:realdata} presents the results of applying the Z-residual diagnostics  tool for diagnosing the functional form of covariates in a real data application. The article is concluded in Section \ref{sec:concl}.

\section{Shared Frailty Model and Statistical Inference}\label{sec:model}
\subsection{Notation and Shared Frailty Model}
A shared frailty model is a frailty model where the frailties are common or shared among individuals within groups. The formulation of a frailty model for clustered failure survival data is defined as follows. Suppose there are $g$ groups of individuals with $n_i$ individuals in the $i$th group, $i$ = 1, 2, \ldots , $g$. If the number of subjects $n_i$ is 1 for all groups, then the univariate frailty model is obtained \citep{KaragrigoriouAlex2011FMiS}. Otherwise, the model is called the shared frailty model \citep{henderson_analysis_2001,duchateau_frailty_2008,hougaard_frailty_1995}  because all subjects in the same cluster share the same frailty value $z_i$. Suppose $t_{ij}$ is the true failure time for the $j$th individual from the $i$th group, which we assume to be a continuous random variable in this article, where $j = 1, 2, \cdots, n_i$. Let $t_{ij}^{*}$ denote the realization of $t_{ij}$. In the scenario of right censoring,  we can observe that $t_{ij}$ is greater than a value $c_{ij}$, where $c_{ij}$ is the corresponding censoring time. The observed failure times are denoted by the pair $(y_{ij}, \delta_{ij})$, where $y_{ij}=\min(t_{ij}, c_{ij}), \delta_{ij}=I (t_{ij} < c_{ij})$. The observed data can be written as $y=(y_{11},\cdots,y_{gn_g})$ and $\delta=(\delta_{11},\cdots,\delta_{gn_g}$).  Since we will consider only the right-censoring in this article, we will use ``censoring'' as a short for ``right-censoring''. The survival function of $t_{ij}$ based on a postulated model is defined as $S_{ij}(\tistar) = P(t_{ij}> t_{ij}^{*})$, where the subscript $ij$ indicates that the probability depends on covariate $x_{ij}$ for the $j$th individual of the $i$th group.

For a shared frailty model, the hazard of an event at time $t$ for the $j$th individual, $j$ = 1, 2, $\cdots$, $n_i$, in the $i$th group, is then
 \begin{equation}
h_{ij}(t) =z_i \exp(x_{ij} \beta)h_0(t);
\end{equation}
and the survival function for the $j$th individual of the $i$th group at time $t$ follows:
 \begin{equation}
S_{ij}(t) = \exp \bigg\{ -  \int_{0}^{t} h_{ij}(t) \, \mathrm{d}t \bigg \}
=\exp \bigg\{-z_i \exp(x_{ij} \beta) H_0(t) \bigg \},
\end{equation}
where $x_{ij}$ is a row vector of values of $p$ explanatory variables for the $j$th individual in the $i$th group, i.e., $x=(x_{11}, \cdots, x_{gn_g})$; $\beta$ is the vector of regression coefficients; $h_0(t)$ is the baseline hazard function, $H_0(t)$ is the baseline  cumulative hazard function (CHF), and $z_i$ is the frailty term that is common for all $n_i$ individuals within the $i$th group. Let $z=(z_1, \cdots, z_g)$. The hazard and survival functions with frailty can also be written as,
\begin{equation}
h_{ij}(t) = \exp(x_{ij} \beta + u_i)h_0(t),
\end{equation}
and
\begin{equation}
S_{ij}(t) = \exp \bigg\{ - \exp(x_{ij} \beta + u_i) H_0(t) \bigg \},
\end{equation}
where $u_i$= $\log (z_i)$ is a random effect in the linear component of the proportional hazards model. Note that $z_i$ cannot be negative, but $u_i$ can be any value. If $u_i$ is zero, corresponding to $z_i$ is one, the model does not have frailty. The form of the baseline hazard function may be assumed to be unspecified as a semi-parametric model or fully specified to follow a parametric distribution.

In our study, we focus mainly on the shared gamma frailty model, since gamma distribution is one of the most common distribution for modelling the frailty effect \citep{collett_modelling_2015}. It is easy to obtain a closed-form representation of the observable survival, cumulative density, and hazard functions due to the simplicity of the Laplace transform \citep{BalanTheodorA2020Atof}. The gamma distribution is a two-parameter distribution with a shape parameter $k$ and scale parameter $\theta$. It takes a variety of shapes as $k$ varies: when $k$ = 1, it is identical to the well-known exponential distribution; when $k$ is large, it takes a bell-shaped form reminiscent of a normal distribution; when $k$ is less than one, it takes exponentially shaped and asymptotic to both the vertical and horizontal axes. Under the assumption $k= \frac{1}{\theta}$, the two-parameter gamma distribution turns into a one-parameter distribution. The expected value is one and the variance is equal to $\theta$.

\subsection{Parameter Estimation and Inference}
Arguably the most popular R package for fitting semi-parametric shared frailty models is the \texttt{survival} package \citep{survival-package}. The \texttt{coxph} function of the \texttt{survival} R package can be used to fit semi-parametric shared frailty models via penalized partial likelihood method \citep{duchateau_penalized_2004,ripatti_estimation_2000-1,mcgilchrist_reml_1993}; and the Breslow (1972) estimator \citep{1972DoPC,lin_breslow_2007} is used for estimating the baseline CHF. The frailty distribution can be specified as Gamma, Gaussian, or t distribution. It accommodates the clustered failures and recurrent events data with the right, left, and interval censoring types. When the \texttt{coxph} function fits the shared frailty model with clustered failures data, the cluster size should be above five. Otherwise, the random effects will be treated as  fixed effects. The \texttt{survival} R package is used for estimating parameters and inference in this study.

In the Cox proportional hazards regression, the Breslow estimator \citep{lin_breslow_2007} is the nonparametric maximum likelihood estimation for the baseline CHF. %It has been implemented in all major statistical software packages. 
The baseline CHF is $H_0(t) = \int_{0}^{t} h_0 (s) \,ds$. Breslow (1972) suggested estimating the baseline CHF via maximizing the likelihood function. After getting the estimators $\hat{\beta}$ and $\hat{u}_i$, nonparametric maximum likelihood estimator of $\hat{H}_0(t)$ can be derived as:
\begin{equation}
\hat{H}_0(t)= \displaystyle\sum_{ \{v: y_{(v)} \le t \} } \bigg  \{ \frac{d_{(v)}} { \displaystyle\sum_{(i, j) \in R_{(y_{(v)})} } \exp(x_{ij} \hat{\beta} + \hat{u}_i )  }  \bigg \},
\end{equation}
where $y_{(1)} < \cdots <  y_{(r)}$ are the ordered distinct event time among the ${y_{ij}}$'s and $R(y_{(v)}) = \{ (i,j): y_{ij} \ge y_{(v)} \}$ is the risk set at $y_{(v)}$, i.e., $d_{(v)}$ is the number of events at $y_{(v)}$. The Breslow approximation is the first option to estimate the baseline hazard function in nearly all the R packages for fitting Cox regression models with or without frailties.

The penalized partial likelihood (PPL) approach can be used to estimate parameters in a shared frailty model \citep{mcgilchrist_reml_1993,duchateau_frailty_2008}. The full data log-likelihood contains the frailty terms $z$, which are assumed to be observed random variables first. The full data log-likelihood follows the joint density of $(y, \delta)$ and $z$, which can be split into two parts. The first part is the conditional likelihood of the data given the frailties, which takes the random effects $u=\log(z)$ as another set of the parameter in the first part of the likelihood. The second part is the log-likelihood of the random effects. Since the full likelihood is only used to estimate the $p$ components of $\beta$ and the $g$ components of $u$, the terms involving $\theta$ alone can be omitted to give the penalized. The second part corresponds to the frailties distribution in which the likelihood is considered a penalty term. The estimation is based on maximizing the penalized partial log-likelihood (PPL) for the frailty model, which is given by 
\begin{equation}\label{ppl} 
l_{ppl}(\beta, u, \theta; y, \delta)= l_{part}(\beta, u; y,\delta) + l_{pen}(\theta; u),
\end{equation}
over both $\beta$ and $u$. Here $l_{part}(\beta,u)$ is the partial log-likelihood for the Cox model that includes the random effects. 
\begin{equation}
l_{part}(\beta, u; y,\delta)=  \displaystyle\sum_{i=1}^{g} \displaystyle\sum_{j=1}^{n_{i}} \delta_{ij} \bigg \{ \eta_{ij} - \log \bigg [ \displaystyle\sum_{(q, w) \in R(y_{ij})} \exp(\eta_{qw}) \bigg ] \bigg \},
\end{equation}
 where $\eta_{ij} = x_{ij}\beta  + u_i $ and $\eta = (\eta_{11},\dots, \eta_{gn_{g}})$.  In the penalty function $l_{pen}(\theta;u)$,  $\theta$ is the parameter for the frailty. The random effect $u$ is equal to $\log(z)$, where $z$ is usually assumed to have a gamma distribution. The penalty function can be written as, 
\begin{equation}
l_{pen}(\theta;u)=  \displaystyle\sum_{i=1}^{g} \log f_{U}({u_{i}|\theta}), 
\end{equation}
where $f_{U}({u_{i}})$ denotes the density function of the random effect $u_i$.

The maximization of the PPL consists of an inner and an outer loop \citep{duchateau_frailty_2008}. For the gamma frailty effects with unit mean and variance $\theta$, the penalized likelihood can be maximized with the Newton-Raphson algorithm in the inner loop. The estimates of $\beta$'s and the $u$'s are first taken to be values that maximize $l_{ppl}(\beta, u, \theta)$ for a given value of  the $\theta$. The outer loop is based on the maximization of a profiled version of the marginal likelihood for $\theta$ given estimates $\hat{\beta}$ and $\hat{u}$. The process is iterated until convergence. %The Breslow approximation is the first option to estimate the baseline hazard function in nearly all the R packages for fitting Cox regression models with or without frailties. (CF: commented out this sentence, since it is repetitive) 

\section{Review of Existing Residuals for Survival Models}\label{sec:review}

In this section, we review some existing residuals used in survival analysis. A central concept in these residuals is formulated based on the survival probability (SP). The widely used CS residual is defined as $r^c_{ij}(t_{ij})=-\log(S_{ij}(t_{ij}))$, where $t_{ij}$ is the true failure time. In the absence of censored observations, the survival probability is uniformly distributed when the model is true; therefore, the CS residual is exponentially distributed. A plot of the CHF against the true failure time will give a straight line through the origin with a unit slope when the residuals have a unit exponential distribution, which is expected when the survival model is correctly specified. In addition to the graphical checking, we can apply numerical GOF testing methods such as Kolmogorov-Smirnov (KS) test to CS residuals. When there are censored failure times, the distribution of $S_{ij}(y_{ij})$ is no longer uniformly distributed under the true model, which means the CS residuals are no longer exponentially distributed. The CS residuals $r^c_{ij}$ can be regarded as a dataset with censoring. The Kaplan-Meier (KM) estimate of the survivor function  can still be computed for CS residuals. Hence, the most widely used diagnostics  tool is to apply the KM method to get an estimate of the CHF of CS residuals and compare the CHF against the $45^o$ straight line. 

Transforming SPs into exponentially-distributed CS residuals is only one option among many others. For example, one can also transform SPs using the quantile of standard normal distribution \citep{nardi_new_1999-1}, defined as $r^{n}_{ij}(y_{ij} )=-\Phi^{-1}(S_{ij}(y_{ij}))$, where $y_{ij}$ is the observed failure time or censoring time. We will call it \textbf{censored Z-residuals} in this paper. The diagnosis of the GOF of $S_{ij}(y_{ij})$ can be converted to the diagnosis of the normality of $r^{n}_{ij}(y_{ij})$. The function \texttt{gofTestCensored} in R package \texttt{EnvStats}  \citep{millard_envstats_2018-1,steven_p_millard_author_envstats_2013} provides an SF test  for testing the normality of multiply censored data. Hence, \texttt{gofTestCensored} can be applied to check the normality of censored Z-residuals for checking the overall GOF of survival models. We will refer to this test using \textbf{CZ-CSF} method in this paper.

Although the aforementioned overall GOF checking methods can be used to determine how closely the residuals are distributed corresponding to their reference distributions when the model assumptions are met, they cannot be used to test the plausibility of specific model assumptions, in particular,  the functional form of covariates. For checking whether a functional form of individual covariates may be misspecified, tailored graphical and quantitative diagnostics tools are needed. Martingale and deviance residuals have been proposed to check the functional form in survival analysis. The martingale residuals \citep{therneau_martingale-based_1990} provide a measure of the discrepancy between the number of predicted death by the model and the number of observed failures in the interval $(0, t_{ij})$, which is either 1 or 0. The martingale residuals are defined as $r^M_{ij}={\delta_{ij}} - r^c_{ij} $, where ${\delta}_{ij}$ is the event indicator for the $j$th individual of the $i$th group observation, ${\delta}_{ij}$ is equal to 1 if that observation is an event; otherwise zero if censored, and  $r^c_{ij} $ is the Cox-Snell residual. The martingale residuals sum to zero, but are not symmetrically distributed about zero \citep{collett_modelling_2015}. The deviance residuals \citep{therneau_modeling_2013,mccullagh_generalized_1989} can be regarded as an attempt to make the martingale residuals symmetrically distributed about zero, and are defined as $ r^D_{ij}= sgn(r^M_{ij}) [-2(r^M_{ij}+{\delta_{ij}}\log({\delta_{ij}}-r^M_{ij}))]^{1\over2}$, where $r^M_{ij}$ is the martingale residual,  the function $sgn(.)$ is the sign function \citep{collett_modelling_2015}. Other residual-based diagnostics  tools have also been proposed for censored survival models; see \citet{residual_mixturecuremodel_2017,grambsch_proportional_1994-1,residual_Tree-Structured_survival,residual_ph_interval_censored,deviance_normal_scores,lin_checking_1993,law_residual_2017,shepherd_probability-scale_2016-1,hillis_residual_1995} and the references therein.  A common drawback for these residuals is that their distributions under the true model are very complicated due to the censoring, hence, they cannot be characterized by  a known distribution or probability table, posing challenges for devising numerical tests based on these conventional residuals for diagnosing survival models.

\section{Z-residual}\label{sec:zresidual}
\subsection{Definition of Z-residual}
In this paper, we extended Z-residual \citep{LiLonghai2021Mdfc}, to diagnose shared frailty models in a Cox proportional hazard setting with a baseline function unspecified. The normalized randomized survival probabilities (RSPs) for $y_{ij}$ in the shared frailty model is defined as:
\begin{equation}
S_{ij}^{R}(y_{ij}, d_{ij}, U_{ij}) =
\left\{
\begin{array}{rl}
S_{ij}(y_{ij}), & \text{if $y_{ij}$ is uncensored, i.e., $d_{ij}=1$,}\\
U_{ij}\,S_{ij}(y_{ij}), & \text{if $y_{ij}$ is censored, i.e., $d_{ij}=0$,} 
\end{array}
\right. \label{rsp} 
\end{equation}
where $U_{ij}$ is a uniform random number on $(0, 1)$, and $S_{ij}(\cdot)$ is the postulated survival function for $t_{ij}$ given $x_{ij}$. $S_{ij}^{R}(y_{ij}, \delta_{ij}, U_{ij})$ is a random number between $0$ and $S_{ij}(y_{ij})$ when $y_{ij}$ is censored. It is proved that the RSPs are uniformly distributed on $(0,1)$ given $x_{i}$ under the true model \citep{LiLonghai2021Mdfc}. Therefore, the RSPs can be transformed into residuals with any desired distribution. We prefer to transform them with the normal quantile:
\begin{equation}
r_{ij}^{Z}(y_{ij}, \delta_{ij}, U_{ij})=-\Phi^{-1} (S_{ij}^R(y_{ij}, \delta_{ij}, U_{ij})),\label{zresid}
\end{equation}
which is normally distributed under the true model, so that we can conduct model diagnostics with Z-residuals for censored data in the same way as conducting model diagnostics for a normal regression model. There are a few advantages of transforming RSPs into Z-residuals. First, the diagnostics methods for checking normal regression are rich in the literature. Second, transforming RSPs into normal deviates facilitates the identification of extremely small and large RSPs. The frequency of such small RSPs may be too small to be highlighted by  the plots of RSPs. However, the presence of such extreme SPs, even very few, is indicative of model mis-specification. Normal transformation can highlight such extreme RSPs.

\subsection{Diagnosis of the Functional Form of Covariates using Z-residuals} \label{sec:testnl}

\begin{figure}[t]
\centering
  \includegraphics[width=2.3in, height=1.9in]{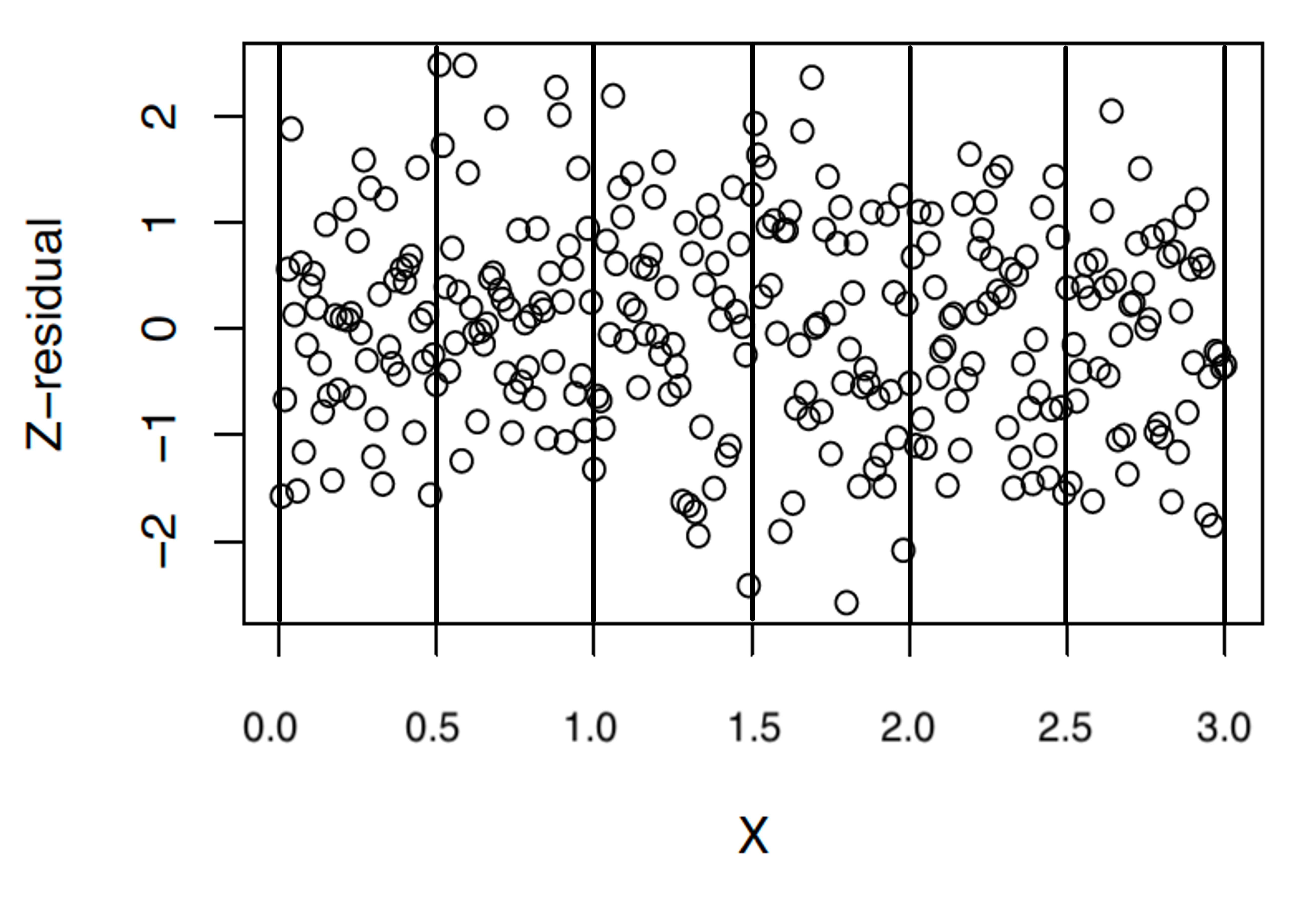} 
  \includegraphics[width=2.3in, height=1.9in]{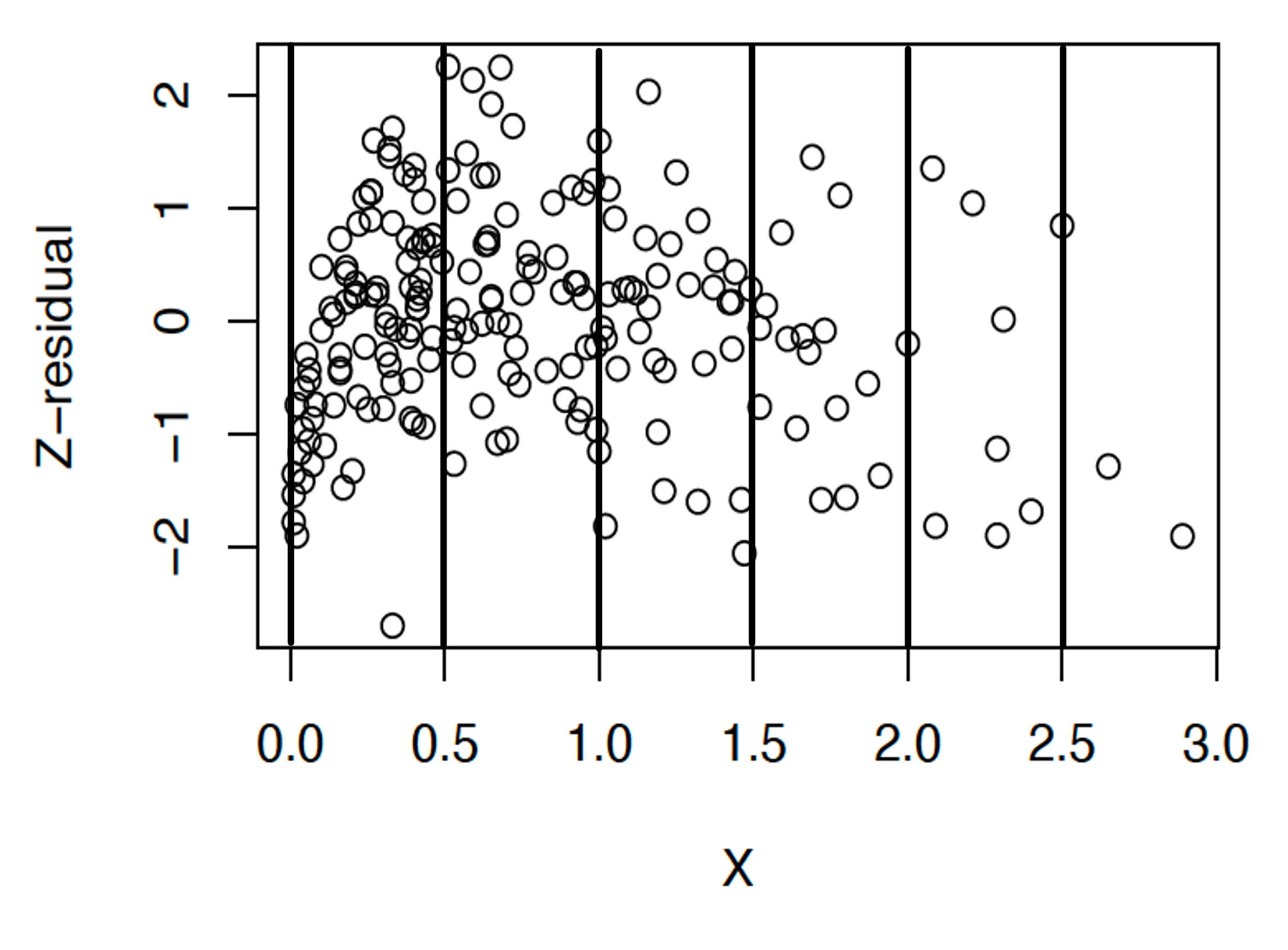} 
%\vspace{-3mm}
\caption{An illustrative plot showing how to construct the non-homogeneity test with Z-residuals: dividing Z-residuals by a covariate or linear predictor (LP) with equally-spaced interval, then testing the equality of the means of  grouped residuals.
 \label{fig:cut_zresidual}}
 \end{figure}

A QQ plot based on Z-residuals can be used to graphically assess the model's overall GOF, and SW or SF test applied to Z-residuals can be used to numerically test the overall GOF of the model. The conditional distribution of Z-residual given $x_i$ is approximately a standard normal and is homogeneous at varying levels of covariates when a model is correctly specified. For checking the functional form of the covariate, we can plot Z-residuals against covariates and/or linear predictors. When the functional form is correctly specified, we expect that there is no trend in these scatterplots. However, such a graphical examination is difficult to determine whether the observed trend in Z-residuals is caused by chance or by the misspecification in the covariate function. Therefore, we desire a formal test to quantify the statistical  significance of the difference between the observed trend and the expected horizontal line at 0. In this paper, we propose the following diagnostics procedure. The Z-residuals can be divided into $k$ groups by cutting the covariates or linear predictors into equally-spaced intervals as shown in Figure  \ref{fig:cut_zresidual}. Then we can check whether the Z-residuals of the $k$ groups are homogeneously distributed. Figure  \ref{fig:cut_zresidual} demonstrates two scatterplots about Z-residuals by cutting the covariate $X$ into equally-spaced intervals. The left panel shows that the Z-residuals are randomly scattered without showing differential  group means or variances. The right panel clearly shows that the Z-residuals are not homogeneous; particularly their group means differ substantially. A quantitative method to assess the homogeneity of such grouped Z-residuals is to test the equality of group means of the Z-residuals. We apply the F-test in ANOVA to  test the equality of the means of grouped Z-residuals as shown in Figure  \ref{fig:cut_zresidual}.

\subsection{A P-value Upper Bound for Assessing Replicated Z-residuals GOF Test p-values}\label{sec:pmin}

A difficulty in conducting statistical tests with Z-residuals is the randomness in the test p-values. Given a fitted model, we can generate many sets of Z-residuals and obtain replicated test p-values. According to the distribution of order statistics of correlated random variables \citep{caraux_bounds_1992,rychlik_stochastically_1992}, we can obtain the following inequality for the $r$th order statistics $p_{(r)}$: 
\begin{equation}
P(p_{(r)}<t) \leq \min\left(1,t{J\over r}\right). 
\label{eqn:pbound}
\end{equation}
Based on \eqref{eqn:pbound}, a p-value upper bound  for observed (simulated) $r$th statistics $p_{(r)}\obs$ is given by  $\min\left(1,p_{(r)}\obs{J\over r}\right)$.  To avoid the selection of $r$, we report the minimal upper bound for $r=1,\ldots, J$, denoted by $\pmin$:
\begin{equation}
\pmin = \min_{r=1,\ldots,J} \min\left(1,p_{(r)}\obs{J\over r}\right).
\end{equation}
The $\pmin$ is rather conservative for assessing model fit because of its generality. When a model has a small $\pmin$, it is highly suspected that the model can be improved for better fitting the dataset. Considering the conservatism of $\pmin$, a rule of thumb for declaring model failure in practice should be much larger, say 0.25 as suggested by \citet{yuan_goodness--fit_2012}, than the conventional $0.05$ for exact p-values.

\section{Simulation Studies}\label{sec:nl}

In this section, we present simulation studies to demonstrate the effectiveness of the Z-residuals in checking the adequacy of the functional form of covariates. Three covariates are generated as follows: $x_{ij}^{(1)}$ is from  Uniform[0, 1], $x_{ij}^{(2)}$ is from positive Normal(0, 1), and  $x_{ij}^{(3)}$ is from Bern(0.25). We generate the failure times $t_{ij}$ from a shared frailty model with Weibull baseline with the following hazard function: 
\begin{equation}
h_{ij}(t_{ij}) =z_i \exp(\beta_1 x_{ij}^{(1)} + \beta_2 \log(x_{ij}^{(2)}) +\beta_3 x_{ij}^{(3)} )h_0(t_{ij}),
\end{equation}
where $h_0$ is the hazard function of Weibull with shape  $\alpha$=3 and scale $\lambda$=0.007. The data generator is given by $t_{ij}= \bigg\{ \frac {- \log(u_{ij})}{\lambda z_i \exp(x_{ij}^{(1)} - 2 \log(x_{ij}^{(2)}) + 0.5 x_{ij}^{(3)} )} \bigg\}^{(1 / \alpha) }$, where $i$ = \{1,$\cdots$, g \} and $j $ = \{1,$\cdots$, $n_i$ \} and $u_{ij}$ is simulated from Uniform ([0, 1]); the frailty term $z_i$ is generated from a gamma distribution with a variance of 0.5. The censoring times $C_{ij}$ are simulated from  exponential distributions. The rates $\gamma$ were set to four different values to obtain four different censoring rates: 0\%, 20\%, 50\%, and 80\%. We fixed the number of clusters $g=20$, and set the cluster size ($n_i$, sample size in each cluster) to be 10 values: 10, 20, \ldots, 100. For each combination of cluster size and censoring rate, we generated 1000 datasets for estimating model rejection rates of different diagnostics methods. In addition to fitting the true model with $\log(x_2)$ as a covariate to these datasets, we also consider fitting the shared frailty gamma model assuming linear effect for $x_2$ as a wrong model to investigate the performance of different diagnostics methods.

\begin{figure}[htbp]
\centering
  \includegraphics[width=5in,height=7.5in]{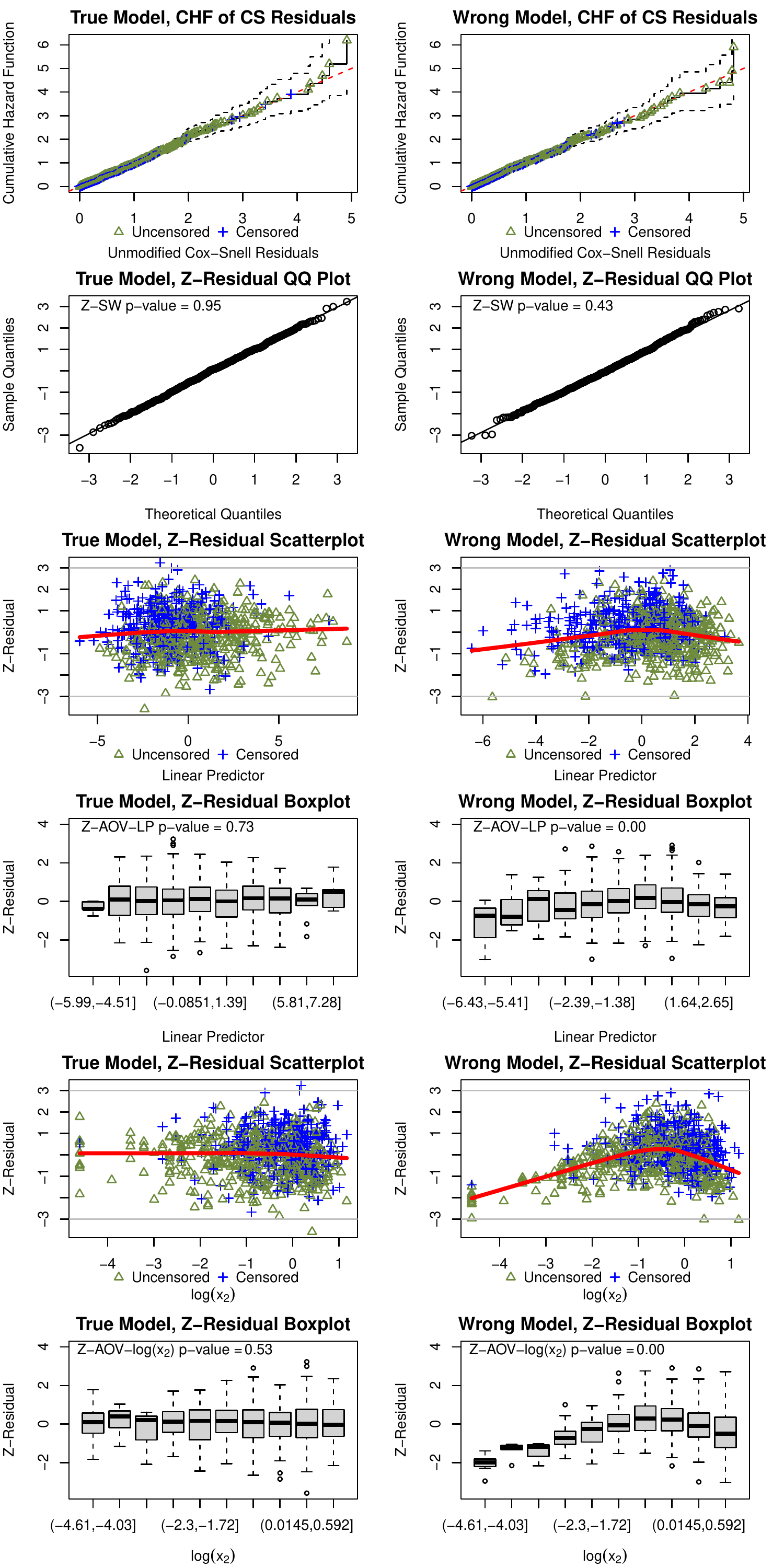} 
%\vspace{-3mm}
\caption{Performance of the Z-residuals and CS residuals as graphical tools for detecting the misspecification of the functional form of covariates. The dataset was generated with 20 clusters of 40 observations in each cluster and a censoring rate $c \approx 50\%$.
 \label{fig: allresid}}
 \end{figure}

We first show the performance of graphical methods for assessing the overall GOF for a single simulated dataset with 20 clusters of 40 observations in each cluster and the percentage of censoring $c\approx 50\%$. As shown in the  panels of the first row of Figure \ref{fig: allresid}, the CHFs of the CS residuals of both of the true and wrong models align well along the $45^{\circ}$ straight line, suggesting that the CS residuals cannot effectively detect the model misspecification of the wrong model with linear covariate effects. The normality of the Z-residual under the true and the wrong models is examined via QQ plots, as shown in the panels of the second row of Figure \ref{fig: allresid}. The points in the two QQ plots for Z-residuals align very well along with straight lines, indicating that the distributions of the Z-residuals under the true and the wrong models are very close to a normal distribution. Therefore, the QQ plots of Z-residuals cannot detect the misspecification in the wrong model either. 

The panels in the third and fourth rows of Figure \ref{fig: allresid} demonstrate the advantage of examining the scatterplots of Z-residuals against the linear predictor for diagnosing the misspecification of the functional form of covariates. Under the true model, the residuals are mostly bounded between -3 and 3 as the standard normal variates without a visible trend. We can see the LOWESS curve in the scatterplot under the true model is very close to the horizontal line at 0. For the wrong model, a non-linear trend in the Z-residuals is clearly observed. In the fourth row, we first divide Z-residuals into $k$ = 10 groups by cutting the linear predictors into equally spaced intervals. The scatterplot and the boxplot indicate that the Z-residuals are homogeneous across groups under the true model, but exhibit differential group means under the wrong model. We further checked the scatterplots and grouped boxplots of Z-residuals against $\log(x_2)$ under the true and wrong models, as shown in the fifth and sixth rows of Figure \ref{fig: allresid}. The Z-residuals of the true models are fairly homogeneous against $\log(x_2)$. By contrast, for the wrong model, we see a clear non-linear pattern in the scatterplots and differential group means in the boxplots against $\log(x_2)$; these plots suggest that the model with linear covariate effects does not fit well to the dataset. 

%The Locally Weighted Scatterplot Smoothing (LOWESS) can help understand the trends of the residual plot.

\begin{figure}[htbp]
\centering
  \includegraphics[width=5in, height=4in]{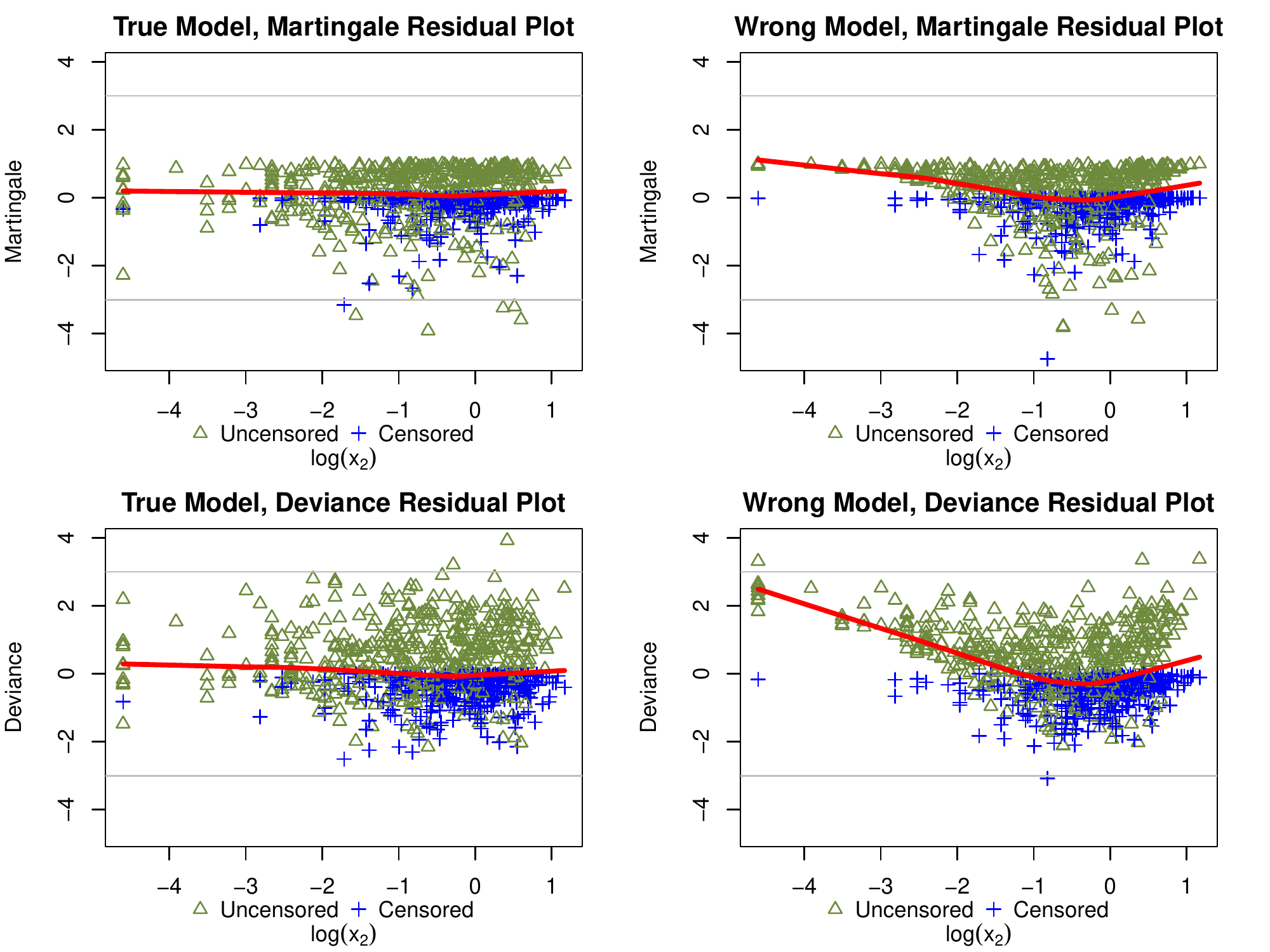} 
\vspace{-3mm}
\caption{Performance of the martingale and deviance residuals as a graphical tool for checking the functional form of covariates. The dataset has a sample size n = 800 and a censoring rate $c \approx 50\%$. 
\label{fig: matg_dev_resid}}
 \end{figure}

As a comparison, we also show the performance of martingale and deviance residuals for assessing the functional form of $x_2$ in Figure  \ref{fig: matg_dev_resid} by displaying the martingale and deviance residuals against the covariate $\log(x_2)$ under the true and wrong models, respectively. Under the true model, the martingale residuals are mostly within the interval (-4, 1); the deviance residuals are more symmetrically distributed than martingale residuals and they are mostly within the interval (-3, 3). The LOWESS curves in the scatterplots of martingale and deviance residuals under the true model are very close to horizontal lines. Note that  the LOWESS curve is slightly tilted downward on the right because  the censoring occurs more frequently for cases with large $\log(x_2)$. Under the wrong model, the LOWESS curves show more pronounced non-horizontal trends in the scatterplots of martingale and deviance residuals. From this comparison, we see that the scatterplots of martingale and deviance residuals can distinguish the true and wrong models and confirm that the true model is a better model for the dataset. However, due to the lack of numerical measures, we cannot tell whether the observed non-horizontal trend is caused by chance or due to misspecified functional form for the covariate. The decision based on visual inspection is often subjective. 

In addition to the graphical assessment, numerical tests with Z-residuals can be constructed as Z-residuals are approximately distributed as the standard normal under the true model.  We compare a set of residual-based testing methods for detecting the inadequacy of fitted models. The overall GOF test methods are denoted by “R-T” with “R” denoting the residual name and “T” denoting the test method. For example, Z-SW is the test method that the normality of Z-residuals is tested with the SW test. In particular, CZ-CSF is the method that the normality of censored Z-residuals (shortened by CZ) is tested by an extended SF method for censored observations, which is implemented with \texttt{gofTestCensored} in the R package \texttt{EnvStats}. For detecting the misspecification in the covariate functional form, we can divide Z-residuals into groups by cutting the linear predictor or a covariate into equally-spaced intervals as shown by the boxplots of Figure \ref{fig: allresid}. We can then test the homogeneity of Z-residuals across the groups. Z-AOV-LP is the method of applying ANOVA to test the equality of the means of Z-residuals against the groups formed with the linear predictor (LP) and Z-AOV-$\log(x_2)$ is the method  of testing the equality of the means of Z-residuals against the groups formed with the covariate $\log(x_2)$. 

\begin{figure}[htbp]
\centering
  \includegraphics[width=6in, height=5in]{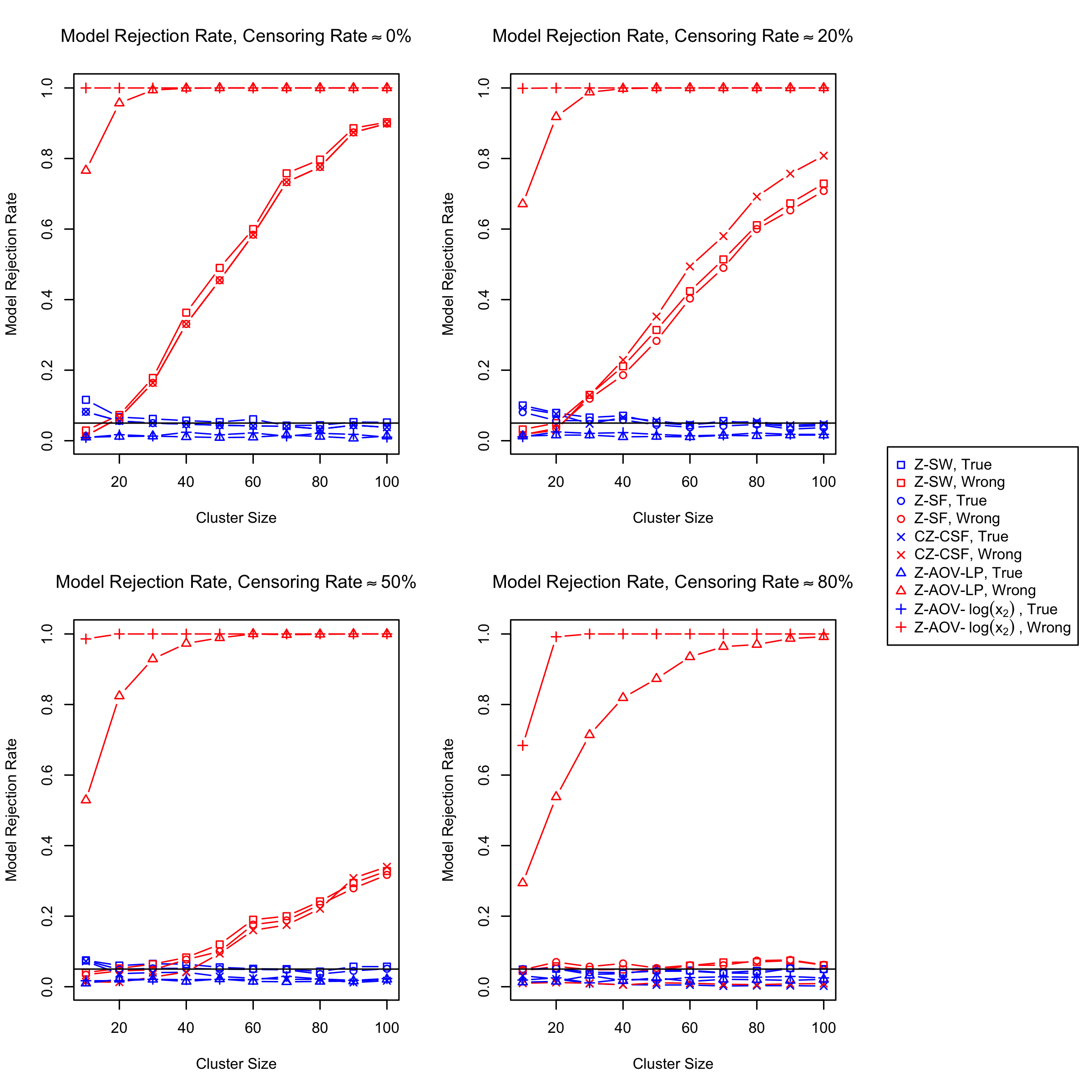} 
\caption{Model rejection rates of various statistical tests based on Z-residual. A model is rejected when the test p-value is smaller than 5\%. Note that we use a random Z-residual test p-value rather than the $\pmin$.}
\label{fig:zresid_gof}
 \end{figure}

We simulated 1000 datasets for each combination of  cluster size and censoring rate as described at the beginning of this section. Using the 1000 datasets generated from the true model under each scenario, the model rejection rate of each test method was estimated by the proportion of the test p-values less than 0.05. The model rejection rates of all the considered test methods are shown in Figures \ref{fig:zresid_gof} and  \ref{fig:devresid_gof}. The non-homogeneity test methods, Z-AOV-LP and Z-AOV-$\log (x_2)$, can detect the non-linear covariate effects with very high true-positive rates (model rejection rates under the wrong models) and low false-positive rates (model rejection rates under the true models). Of all the compared test methods, Z-AOV-$\log (x_2)$ performs the best for detecting the nonlinear covariate effects with the highest powers, which are nearly 100\%, and the powers stay high even for the scenario with a cluster size as small as 10. The Z-SW, Z-SF, and CZ-CSF tests have false-positive rates close to the nominal level of 5\% for all scenarios and have certain powers when the censoring rate is less than 80\%. We also note that their powers increase as the cluster size increases.  However, the powers of these overall GOF tests are significantly smaller than the corresponding powers of the Z-AOV-LP and Z-AOV-$\log (x_2)$ methods. The comparison demonstrates the advantage of testing the homogeneity of Z-residuals for checking the assumption of covariate functional form in addition to the overall GOF tests, which do not inspect the relationship between residuals and covariates.  

In appended Figure \ref{fig:devresid_gof}, we show the performances of the Z-KS and Dev-SW tests, which were separated from Figure \ref{fig:zresid_gof} for better visualization. Z-KS test has low false-positive rates but also very low powers, which shows the conservatism of the KS test for testing the normality of Z-residuals. When the censorship is 0, the performance of Dev-SW is satisfactory. However, when there are censored observations, the Dev-SW method has very high (nearly 100\%) model rejection rates when the model is correctly specified. Hence, the high powers of Dev-SW do not indicate that it is a good test method.

\section{A Real Data Example}\label{sec:realdata}
In this section, we apply the proposed residual diagnostics tools based on Z-residuals to diagnose the functional form of covariates in a real application for modelling the survival times of acute myeloid leukemia patients. The dataset contains 1498 patients recorded at the M. D. Anderson Cancer Center between 1980 and 1996 \citep{ESTEYE.H2000Eott}. The dataset used in our analysis contains 411 patients who are aged below 60 from 24 administrative districts recorded at the M.D Anderson Cancer Center between 1980 and 1996. The data collected information on the survival time for acute myeloid leukemia and prognostic factors, including  age, sex, white blood cell count (wbc) at diagnosis, and the townsend score (tpi) for which higher values indicate less affluent areas. The censoring rate is 29.2\%. The response variable of interest is the survival time in days, which is the time from entry to the study or  death. The preliminary study showed that the wbc is highly right-skewed. Logarithm transformation is often used to reduce the impact of extremely large values of the covariate on the response variable, such as the wbc variable in this application. However, a logarithm transformation may mask the impact of extremely large values of the covariate on the outcome variable.

We fitted two shared frailty models, one with covariates wbc, age, sex and tpi, which is labelled as the wbc model, and the other with log(wbc) replacing wbc, which is labelled as the lwbc model. Table \ref{tab:covariates} shows the estimated regression coefficients, the corresponding standard errors and p-values for the covariate effects from fitting the two shared frailty models. The results indicate that the estimated effect of wbc is statistically significant (p-value $<$ 0.001) but the effect of log(wbc) is not significant (p-value=0.135). The difference in the p-values for wbc and log(wbc) highlights that the statistical inference of the covariate effect may depend on the assumption of the functional form of the covariates.

\begin{table}[htb]
    \caption{Parameter estimates of the shared gamma frailty model in the real data application. \label{tab:covariates}}
    \normalsize
    \begin{subtable}{.5\linewidth}
      \centering
%       \normalsize
        \caption{ The wbc model\label{tab:covariates1}}
          \begin{tabular}{| l |l | l | l |}
\hline
Covariates&Estimate &SE & P-value \\
\hline
$Age $ & 0.021 & 0.005 & 0.000  \\
\hline
$SexMale$ & 0.215 &0.118&0.068  \\
\hline
$wbc$ &0.005 &   0.001  &0.000 \\
\hline
$tpi$ & 0.023&  0.016&0.140\\
\hline
$Frailty$ &  &  &0.906 \\
\hline
\end{tabular}
   \end{subtable}%
   \hfill
      \begin{subtable}{.5\linewidth}
      \centering
%        \small
        \caption{ The lwbc model \label{tab:covariates2}}
        \begin{tabular}{| l |l | l | l |}
        \hline
Covariates&Estimate &SE & P-value \\
\hline
$Age $ & 0.021 &0.005 & 0.000  \\
\hline
$SexMale$ & 0.216 &0.118&0.069  \\
\hline
$log(wbc)$ &0.035 &0.024 &0.135 \\
\hline
$tpi$ &0.024 &0.016&0.128\\
\hline
$Frailty$ &  &  &0.906 \\
\hline
\end{tabular}
   \end{subtable} 
\end{table}

\begin{figure}[htbp]
\centering
  \includegraphics[width=5in, height=7.5in]{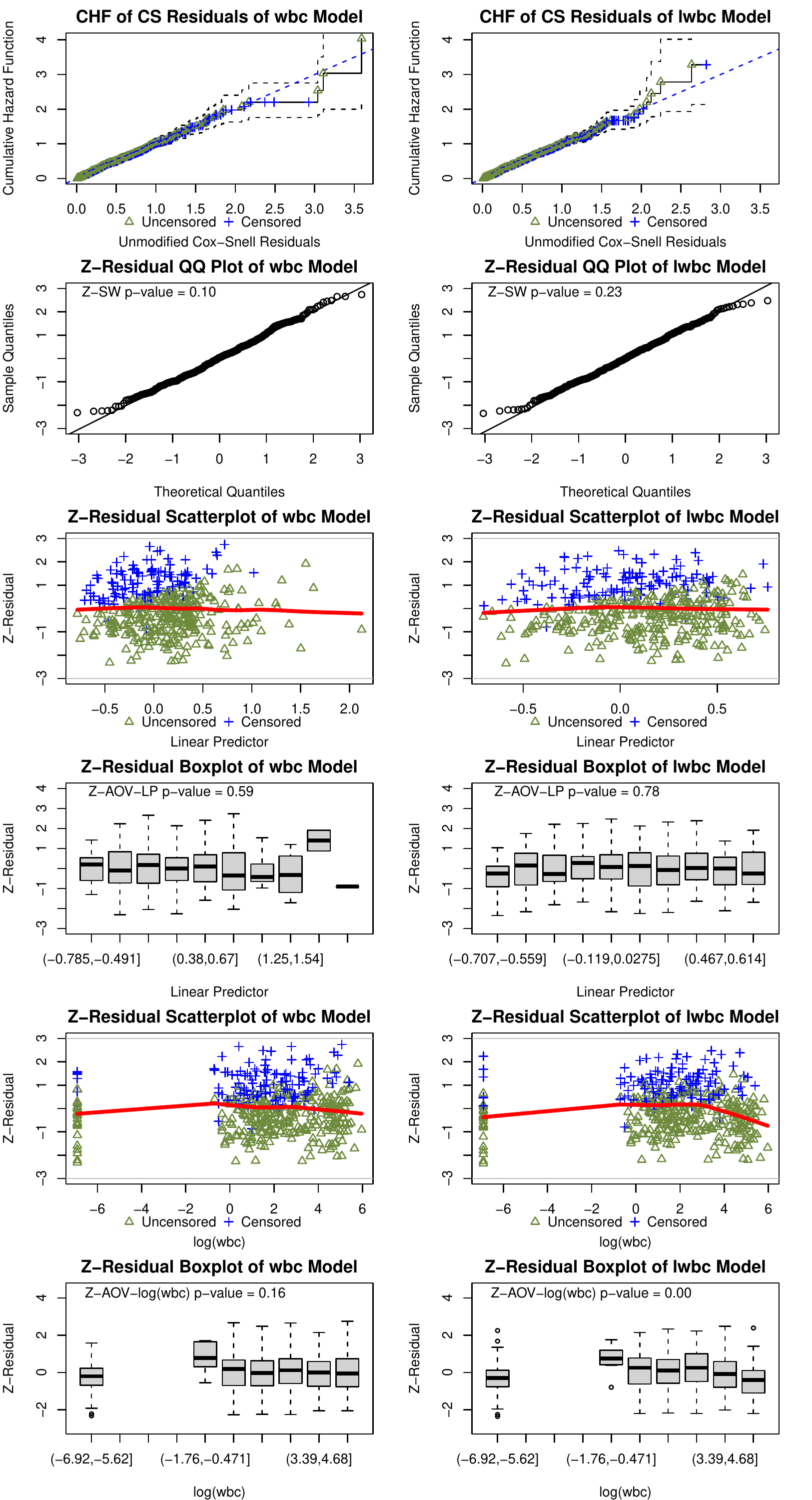} 
%\vspace{-3mm}
\caption{Diagnostics results for the wbc (left panels) and lwbc (right panels) models  fitted to the survival data of acute myeloid leukemia patients.}
 \label{fig: leuksurv_logwbc}
 \end{figure}

The overall GOF tests and graphical checking with CS residuals and Z-residuals show that both the wbc and lwbc models provide adequate fits to the dataset. The first row of Figure  \ref{fig: leuksurv_logwbc} shows that the estimated CHFs of the  CS residuals of both of the wbc and lwbc models align closely along the $45^{\circ}$ diagonal line.  Similarly, the QQ plots (the second row of Figure \ref{fig: leuksurv_logwbc}) of Z-residuals of these two models align well with the 45$^\circ$ diagonal line.  The scatterplots of Z-residuals against the linear predictor don't exhibit visible trends; their LOWESS lines are very close to the horizontal line at 0; the boxplots of Z-residuals grouped by cutting  linear predictors into equal-spaced intervals  (the fourth row of Figure \ref{fig: leuksurv_logwbc}) appear to have equal means and variance across groups. The Z-AOV-LP test also gives large p-values   for the wbc and lwbc models (0.63 and 0.76  respectively).  

The above diagnostics results reveal no serious misspecification in  these two models.  However, the inspection of the Z-residuals against the covariate log(wbc) reveals that the functional form of the lwbc model  is likely misspecified. The scatterplots and comparative boxplots of the Z-residuals against log(wbc) are shown in the fifth and sixth rows of Figure \ref{fig: leuksurv_logwbc}. The LOWESS curve of  the wbc model appears to align well with the horizontal line at 0 and the grouped Z-residuals of the wbc model appear to have equal means and variances across groups. However, the diagnosis results  for the lwbc model are very different. It appears that there is a non-linear trend in the LOWESS curve of the lwbc model and the grouped Z-residuals appear to have different means across groups.  To measure the statistical significance of the observed trends, we apply Z-AOV-$\log(\mbox{wbc})$  to test the equality of the means of the grouped Z-residuals of these two models. The  p-values are 0.16 and 0.00 respectively for the wbc and lwbc models as shown in the boxplots. The very small p-value of the Z-AOV-log(wbc) test for the lwbc models  strongly suggests that the log transformation of wbc is likely inappropriate for modelling the survival time.

\begin{figure}[htpb]
\centering
  \includegraphics[width=5in, height=6.9in]{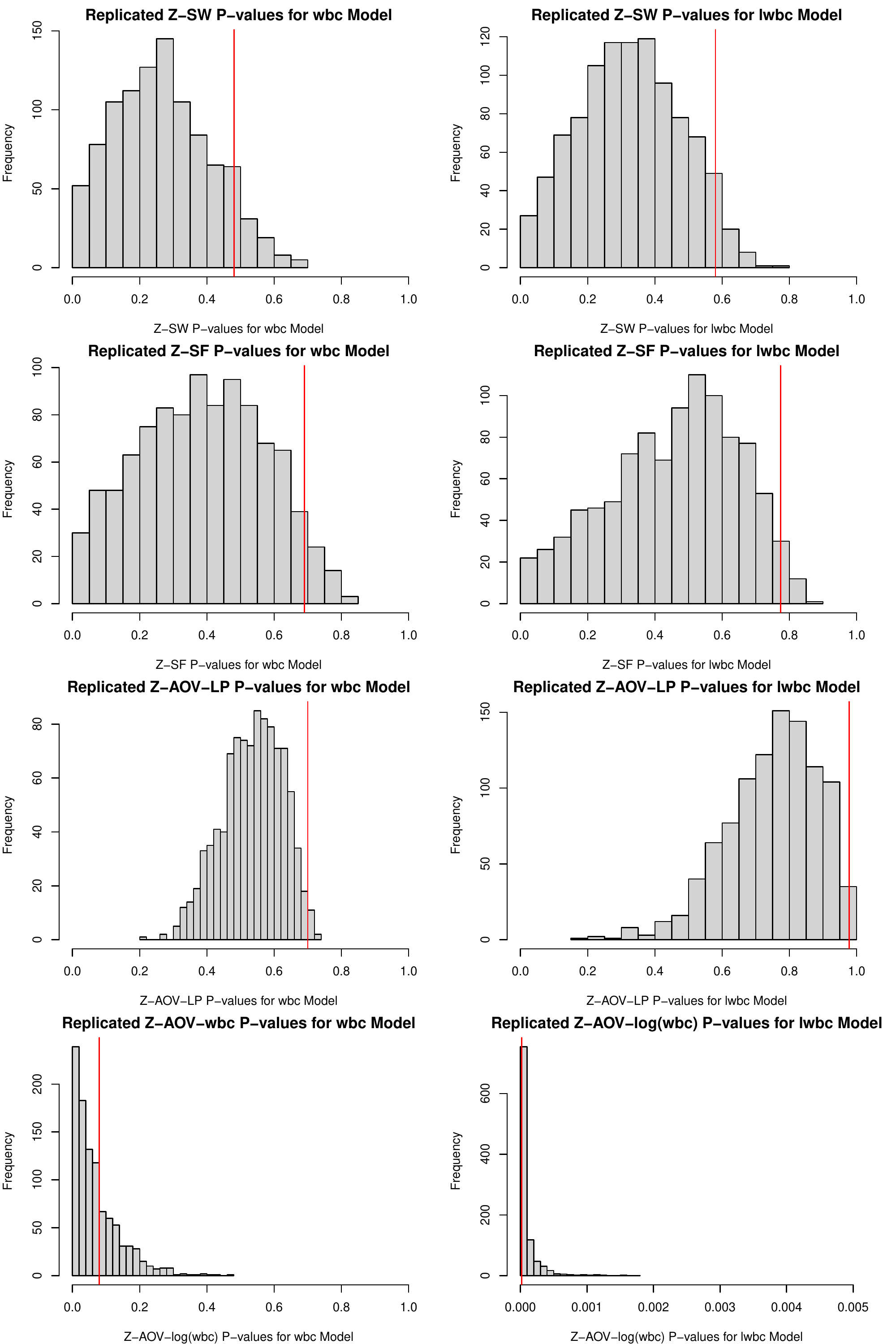} 
\vspace{-3mm}
\caption{The histograms of 1000 replicated Z-SW, Z-SF, Z-AOV-LP and Z-AOV-log(wbc) p-values for the wbc model (left panels) and the lwbc model (right panels) fitted with the survival times of acute myeloid leukemia patients. The vertical red lines indicate $\pmin$ for 1000 replicated p-values. Note that the upper limit  of the x-axis for Z-AOV-log(wbc) p-values for the lwbc model is 0.005, not 1 for others.}\label{fig:hist_Leuk}
\end{figure}

The Z-residual test p-values quoted above contain randomness because of the randomization in generating Z-residuals. To ensure the robustness of the model diagnostics results, we generated 1000 replicated test p-values with 1000 sets of regenerated Z-residuals for each test method. Figure \ref{fig:hist_Leuk} displays the histograms of 1000 replicated Z-residual test p-values for the wbc and lwbc models. The red vertical lines in these histograms show the upper bound summaries of these replicated p-values, $\pmin$ (see Sec. \ref{sec:pmin} for details). These histograms show that the Z-SW, Z-SF, and Z-AOV-LP tests for both models give a large proportion of p-values greater than 0.05, and the large p-values result in large $\pmin$ values.  In contrast, the replicated Z-AOV-log(wbc) p-values for the lwbc model are almost all smaller than 0.001. The consistently small Z-AOV-log(wbc) p-values further confirm that the log transformation of wbc is inappropriate for modelling the survival time.

Table \ref{tab:realdata} tabulates all the $\pmin$ values (shown with red lines in Figure  \ref{fig:hist_Leuk}) for diagnosing the two models with Z-residual-based tests. In addition, we also report the non-random CZ-CSF test p-values for the two models and the AIC values for comparing these two models. The CZ-CSF p-values of both models are larger than 5\% (Table \ref{tab:realdata}). Therefore, the CZ-CSF test does not identify the inadequacy of the lwbc model either. The AIC value, 3132.105, of the lwbc model, is much larger than the AIC value 3111.669 of the wbc model, which indicates that the wbc model provides a better model fit compared to the lwbc model. This conclusion is consistent with the model diagnostics results as given by the Z-AOV-log(wbc) test, which reveals that the lwbc model is inappropriate for modelling the survival time of this dataset by checking the homogeneity of Z-residuals against log(wbc).  Although the AIC of the wbc model is smaller than that of the lwbc model, we also see that a large proportion of Z-AOV-log(wbc) p-values for the wbc model are tiny; the $\pmin$ value is 0.074. We think that the wbc model could be improved to provide a better fit for the survival time of this dataset. 
 
 \begin{table}
\centering
\caption{AIC, p-values or $\pmin$ values for the CZ-CSF test, $\pmin$ for Z-SW, Z-SF, Z-AOV-LP and Z-AOV-log(wbc) test for the wbc and lwbc models, respectively, for the acute myeloid leukemia data. \label{tab:realdata}}
%\setcellgapes{5pt}
%\makegapedcells
\begin{tabular}{| l | l | l | l | l | l | l |}
\hline
Model&AIC & CZ-CSF  & Z-SW &Z-SF &Z-AOV-LP  &Z-AOV-log(wbc) \vspace{-0.02in}\\ 
& & $p-$value &$\pmin$ &$\pmin$ &$\pmin$ &$\pmin$\\ 
\hline
 wbc model&3111.669  &0.255 & 0.495  &0.693  &0.703 & 0.074\\
\hline
lwbc model &3132.105 &0.305  &0.579 &0.781 &0.978 & $<$\textbf{0.00001}   \\
\hline

\end{tabular}
\end{table}
 
%\vspace*{-15pt}
\section{Conclusions and Discussions}\label{sec:concl}

In this paper, we extended the idea of randomized survival probability \citep{LiLonghai2021Mdfc} to develop a residual diagnostic tool that can provide both graphical and numerical results for checking the covariate functional form in semi-parametric shared frailty models. We  proposed a non-homogeneity test for testing whether there is a trend in Z-residuals for checking the covariate functional form. Our extensive simulation studies showed that the overall GOF tests (including CS-CSF, Z-SW, and Z-SF) may not be powerful enough for detecting the misspecification in covariate functional form and that the proposed non-homogeneity tests based on the Z-residuals are significantly more powerful than the aforementioned overall GOF tests. Applied to  a real dataset, the Z-residual diagnostics discovers that a model with log-transformation is inappropriate for modelling the survival time of acute myeloid leukemia patients, which is not captured by other diagnostics methods.

The Z-residuals-based diagnostics methods can be extended in several directions. When the full dataset is used to estimate the model parameters and used to calculate residuals for model checking, there might be a conservatism problem (bias) due to the double use of the dataset. The double use of the data may reduce the power of detecting model misspecification, especially in the case of small sample size or high censoring rate. Cross-validation could be a good method to solve this problem. The cross-validatory Z-residual may be a more powerful tool for identifying the model inadequacy in the survival data.

In this paper, we considered semiparametric shared frailty models assuming proportional hazards. However, if the model includes time-varying coefficients or time-dependent explanatory variables, the proportional hazards assumption may be violated. A number of residuals have been proposed for evaluating the assumption of proportional hazards.  Traditionally, the Schoenfeld \citep{collett_modelling_2015,SCHOENFELDDAVID1982Prft} and Scaled Schoenfeld \citep{grambsch_proportional_1994-1} residuals are often used in testing the assumption of proportional hazard. \citet{lin_checking_1993} proposed the cumulative sums of martingale residuals to check the validity of the PH assumption. Extending the Z-residual for diagnosing the proportional hazard assumption and comparing it with existing residual diagnostics tools warrants a research topic in the future.

\newpage

\appendix

%\section*{Supplementary Materials}

\section{Additional Figures and Tables}

 \begin{figure}[htbp]
\centering
  \includegraphics[width=6in, height=5in]{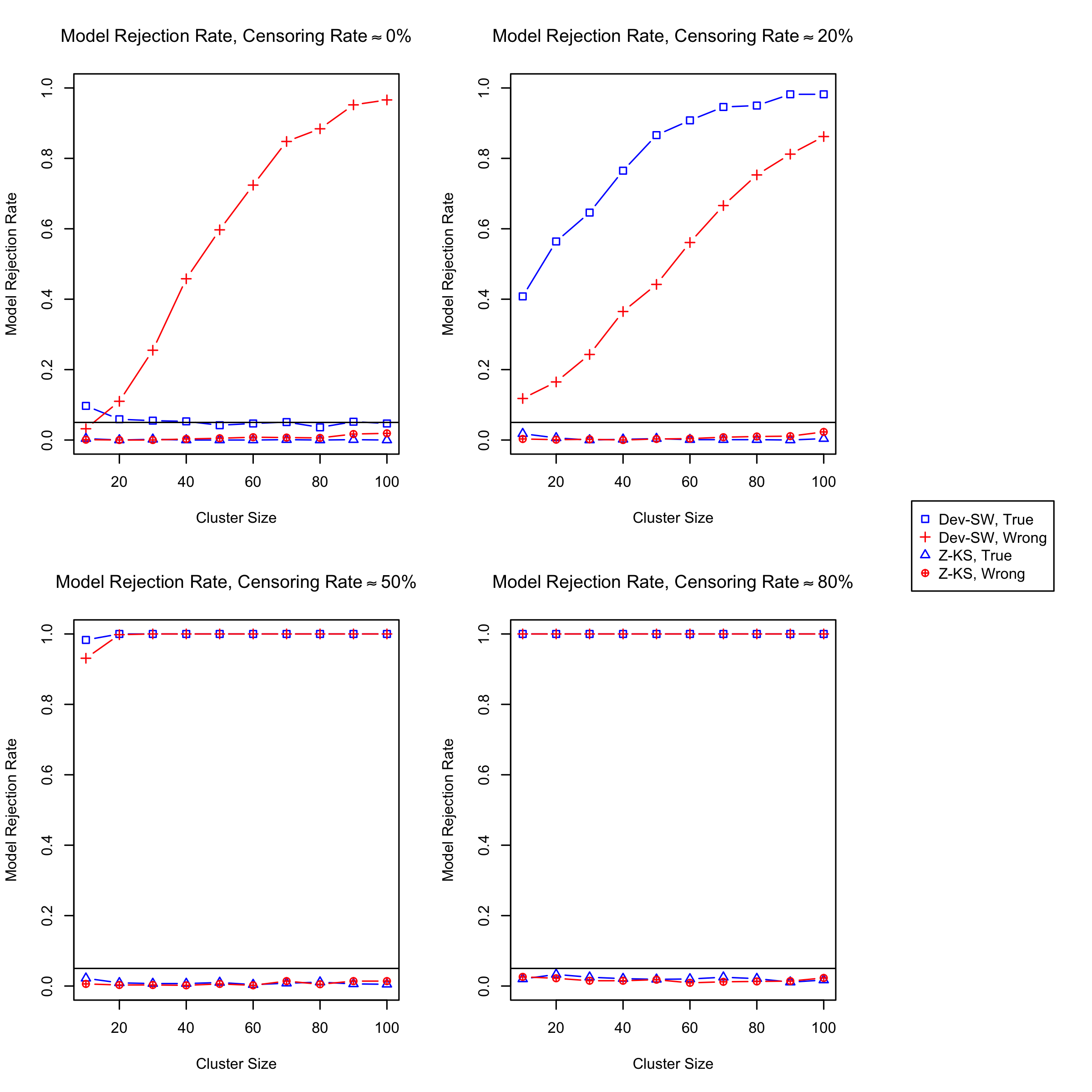} 
\caption{Model rejection rate of the KS test applied to Z-residuals (Z-KS) and the SW test applied to deviance residuals (Dev-SW) for the simulation study in Sec. \ref{sec:testnl}. A model is rejected when the test p-value is smaller than 5\%. The model rejection rates of Dev-SW tests are nearly 1 under the true and wrong models when the censoring rate is 50\% and 80\%, hence, they are almost overlapped in the plots. }
 \label{fig:devresid_gof}
 \end{figure}
 
%\section*{Acknowledgements} 
% The authors would like to acknowledge the support fromthe  Natural Sciences and Engineering Research Council of Canada (NSERC).

%\begin{acks}
%This class file was developed by Sunrise Setting Ltd,
%Brixham, Devon, UK.\\
%Website: \url{http://www.sunrise-%setting.co.uk}
%\end{acks}

%\begin{thebibliography}{99}
\bibliographystyle{Sageh}
\bibliography{z-residual}

%\end{thebibliography}

\end{document}